\documentclass[particles,article,accept,pdftex,moreauthors]{Definitions/mdpi} 
\firstpage{1} 
\pubvolume{1}
\issuenum{1}
\articlenumber{0}
\pubyear{2025}
\copyrightyear{2025}
\datereceived{ } 
\daterevised{ } 
\dateaccepted{ } 
\datepublished{ } 
\hreflink{https://doi.org/} 

\Title{Search for Primordial Black Holes from the International Space Station with the SQM-ISS detector}

\TitleCitation{Primordial Black holes from the ISS}

\Author{Francesca Liberatori $^{1}$\orcidA{}, Matteo Battisti, Marco Casolino$^{1,2}$,   Laura Marcelli $^{2}$, Zbigniew Plebaniak$^{2}$, Enzo Reali$^{1}$}

\AuthorNames{Francesca Liberatori, Matteo Battisti, Marco Casolino, Laura Marcelli, Zbigniew Plebaniak and Enzo Reali }

\isAPAStyle{%
       \AuthorCitation{Liberatori, F., et al.}  }{%
        \isChicagoStyle{%
        \AuthorCitation{Liberatori, Francesca, et al.}
        }{
        \AuthorCitation{Liberatori, F., et al.}
        }
}

\address{%
$^{1}$ \quad Department of Physics, University of Rome Tor Vergata, Via della Ricerca Scientifica 1, 00133 Rome, Italy; email: francesca.liberatori@roma2.infn.it \\
$^{2}$ \quad INFN Structure of Rome Tor Vergata, Via della Ricerca Scientifica 1, 00133 Rome, Italy}

\corres{Correspondence: francesca.liberatori@roma2.infn.it; }

\abstract{
In this paper we discuss the observational capabilities and sensitivity of the SQM-ISS detector to primordial black holes.
The SQM-ISS experiment aims to detect slow, non relativistic massive particles within cosmic rays, using a  detector on board the International Space Station. 
The device is designed to recognize the passage of highly penetrating and dense particles in a wide range of mass and charge states such as Strange Quark Matter (SQM).
These particles, traveling at speeds typical of gravitationally bound objects in the galaxy - around 250 km/s - are also possible candidates  of dark matter.
The ability of SQM-ISS to identify penetrating, massive and slow-moving objects allows it also to be sensitive to the detection of  primordial black holes.
We  discuss how  black holes, traveling through the detector at velocities compatible with galactic orbital speeds, can be identified based on their interaction signatures.
}


\begin{document}



\section{Introduction}

The SQM-ISS experiment primary goal is to look for slow, non-relativistic massive particles within cosmic rays, using a detector on board the International Space Station. 
The device is designed to recognize the passage of highly penetrating and dense particles in a wide range of mass and charge states~\cite{3__PhysRevD.9.3471, 4___PhysRevLett.52.1265}.
One particularly interesting candidate of this kind of particles is Strange Quark Matter (SQM), which is composed of aggregates of up, down, and strange quarks and could represent the fundamental state of hadronic matter~\cite{1___PhysRevD.30.272}. 
These particles, traveling at speeds typical of gravitationally bound objects in the galaxy - around 250 km/s - are also possible candidates of dark matter~\cite{8__PhysRevD.90.045010}.
 The detector can recognize the characteristic signals of these particles by measuring their speed and distance traveled using four plastic scintillator layers read by silicon photomultipliers. In addition, four interleaved metal plates equipped with piezoelectric sensors detect mechanical vibrations caused by the passage of particles.
The time-of-flight system determines the velocity of particles by recording signals through segmented planes of scintillators. The system can detect objects with travel times up to several 2.5 $\mu$s, allowing them to be distinguished from relativistic cosmic rays.The electronics integrate fast analogue-to-digital and time-to-digital converters, with a programmable logic trigger to select relevant events. 
The ability of SQM-ISS to identify penetrating, massive and slow-moving objects allows it also to be sensitive to the detection of Primordial black holes(PBHs)~\cite{SQM-ISS}.

Primordial black holes are thought to have formed in the early Universe due to the collapse of density fluctuations upon reentry into the cosmological horizon during inflationary scenarios~\cite{CarrHawking1974}.
Alternative mechanisms include the role of cosmic strings, domain walls or phase transitions\cite{Chapline1975}, which naturally induce peaks in the PBH mass distribution~\cite{Green2023PBH}.
If black holes evaporate by emitting Hawking radiation, the surviving ones should have a mass greater than $10^{11}$ g. However, string theory suggests that primordial black holes could be interpreted as fuzzballs, without a singularity at its centre, and without a clear horizon. In this case they would not evaporate like traditional black holes. Under this hypothesis, PBH of the fuzzball type  could be long-lived and contribute to dark matter, partially avoiding evaporation via Hawking radiation~\cite{Mathur2017}.
Their detection would provide insights into the conditions of the early universe, inflationary scenarios, and could help constrain their possible role as a component of dark matter. 
SQM-ISS will explore a range of relatively low-mass PBHs ($10^8 - 10^{15}$ g), providing new constraints on their abundance and potential connection to dark matter.

\section{The SQM-ISS project}\label{det}

The SQM-ISS project is a space-based experiment aimed at detecting massive, non-relativistic particles potentially present in cosmic rays. These include exotic candidates such as Strange Quark Matter (SQM), Q-balls~\cite{3___PhysRevLett.80.3185}, compact fermionic objects~\cite{5___PhysRevD.74.063003}, primordial black holes, mirror matter~\cite{8___FOOT2000171}, and Fermi balls~\cite{9___arxiv.hep-ph_9412264}. The project was submitted in response to the 2022 ESA Call for Ideas under the "Reserve Pools of Science Activities for ISS: A SciSpacE Announcement of Opportunity" and selected the following year. 

The instrument will be installed onboard the International Space Station (ISS), which provides a unique observational environment: its low Earth orbit allows direct exposure to cosmic radiation without atmospheric attenuation, enabling sensitivity to lower-mass particles that would otherwise be blocked or slowed down by the Earth's atmosphere. Moreover, the microgravity environment and the absence of terrestrial seismic noise make it ideal for detecting extremely rare and weakly interacting events.
These hypothetical particles, presumed to be of interstellar origin, are expected to be gravitationally bound to the galactic halo and to travel at velocities typical of galactic rotation, approximately $220$–$250\,\mathrm{km/s}$~\cite{SQM-ISS}.

\begin{figure}[H]
\includegraphics[width=0.99\textwidth]{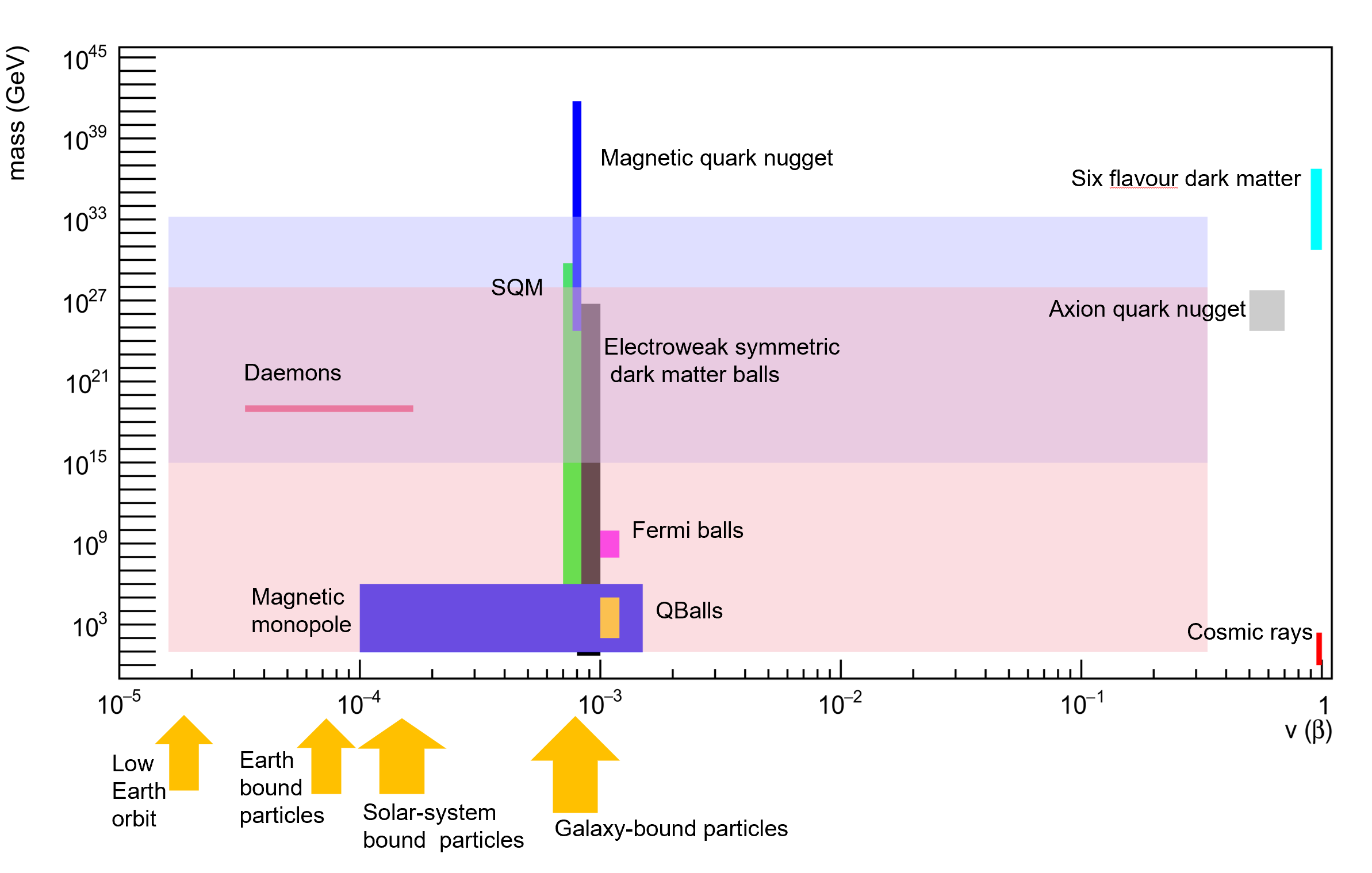}
\caption{Plot of velocity
 and mass range for various hypothetical slow moving, massive particles. At~the bottom right of the plot, we have cosmic ray nuclei with $\beta\simeq 1$ (lower energy particles would be stopped by the hull of the ISS) and mass $\lesssim 200\: $GeV. Most of these hypothetical  particle types have $\beta \simeq 7\text{--}8\times 10^{-4}$ since they are usually considered to be bound in our galaxy. However many of these  particles can also have lower velocity since they could be bound in the  local interstellar medium,  in~the vicinity of the solar system or even of our planet (e.g., Daemons)~\cite{bernabei2018}. The~masses range from the relatively light magnetic monopoles and Q-Balls to the heavier Fermi and dark matter balls. SQM and magnetic quark nuggets can reach the mass of a star, but~only lighter fragments  are expected to reach Earth. The~expected ranges for velocities, masses and mass densities of these different candidates remain hypothetical with wide ranges acceptable for  these values.   This variety and uncertainties are an additional motivation to design an instrument sensitive to a wide range of masses, using complementary means of detection. In~light pink, we show the mass--velocity range accessible by the scintillator/SiPM detectors and in light blue, the range of the piezoelectric detectors. The~ main target  are thus very dense particles that would  move at typical galactic orbital velocities, in~the velocity range between $3\times 10^{-5}$ and $3\times  10^{-1}$ c.}
\label{xxzoo}        
\end{figure}

\begin{figure}[H]
    \centering
    \includegraphics[width=0.95\textwidth]{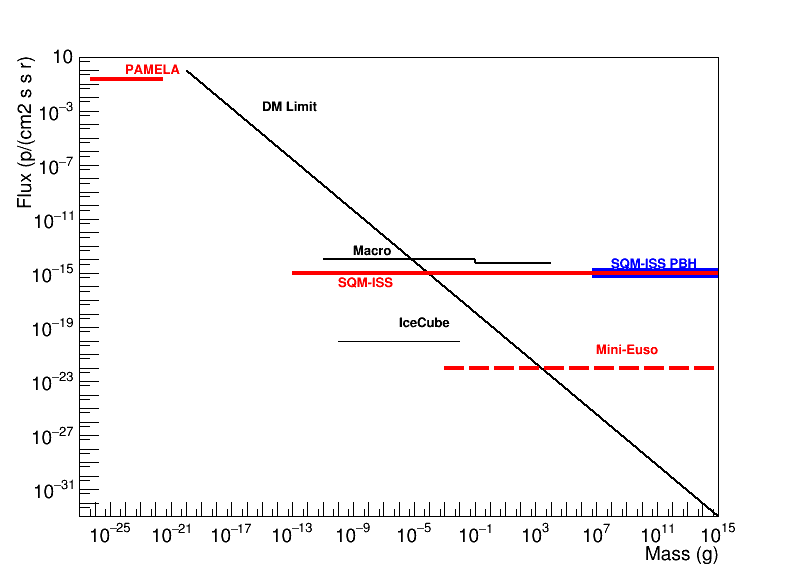}
    \caption{Projected sensitivity of the SQM-ISS experiment compared to current upper limits on the flux of exotic compact particles as a function of mass \cite{SQM-ISS}.}
    \label{fig:sqm_sensitivity}
\end{figure}

As shown in Figure~\ref{xxzoo}, the SQM-ISS detector is designed to probe a wide region of parameter space in mass and flux for a variety of exotic compact particle candidates. These objects are expected to be significantly more massive than atomic nuclei and to exhibit high penetrating power due to their large mass-to-charge ratios~\cite{HaenselChapter8}.
The experiment also aims to be sensitive to particles with lower velocities ($\lesssim 90\,\mathrm{km/s}$), which may be trapped in long-lived or Earth-crossing orbits. These include Strongly Elongated Earth-Crossing Heliocentric Orbits (SEECHOs), Near-Earth Almost Circular Heliocentric Orbits (NEACHOs), and Geocentric Earth Surface-Crossing Orbits (GESCOs). The internal placement of the detector within the ISS provides shielding from external radiation while maintaining the ability to detect highly penetrating slow-moving objects.

\section{Detector Description}
The SQM-ISS detector is specifically designed to identify dense and massive particles through a combination of mechanical and ionization signals, complemented by time-of-flight (ToF) measurements. The detection system integrates scintillators and piezoelectric sensors, allowing the simultaneous acquisition of energy deposition and mechanical impulses, together providing a distinctive signature for such exotic particles.
By correlating ionization and mechanical impact data with precise ToF timing, the detector can effectively discriminate between candidate events and background noise.
The detector is composed of a vertical stack of eight detection planes, each with dimensions of $10 \times 10 \times 0.5$ cm³. The planes are arranged in an alternating sequence of two distinct types: four plastic scintillator layers for ionization-based detection and four copper layers  for detection of mechanical vibrations.
The overall detector height is ~7cm.  

\begin{figure}[H]
    \centering
    \includegraphics[width=0.8\textwidth]{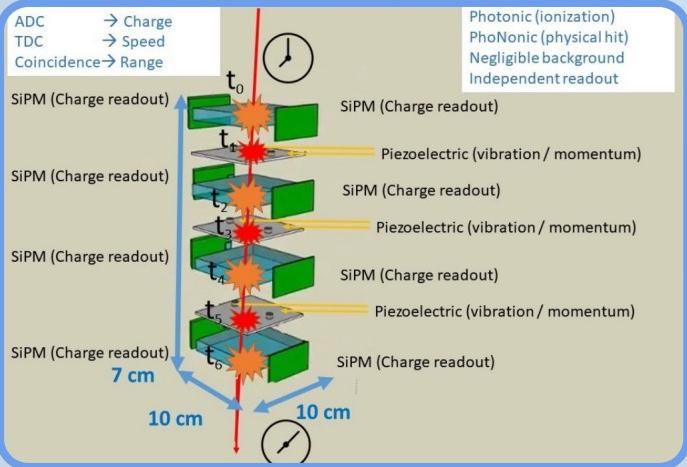}
    \caption{Conceptual scheme of the SQM-ISS detector. The apparatus consist of an alternating stack of plastic scintillator planes and piezoelectric sensor, optimized for detection of slow-moving, penetrating particles such as Strange Quark Matter (SQM). Photonic signals generated by ionization in the scintillators are read by Silicon Photomultipliers (SIPMs), providing charge measurements (via ADC) and time-of-flight information (via TDC) accross multiple layers. Piezoelectric plates detect local mechanical vibrations caused by particles impact,providing an independent phononic signal correlated with momentum transfer. The combination of velocity, charge, and penetration depth allows for the effective identification of SQM against backgrounds from low energy protons and cosmic rays}
    \label{fig:scheme_detector}
\end{figure}
   
\begin{figure}[H]
    \centering
    \includegraphics[width=1.0\linewidth]{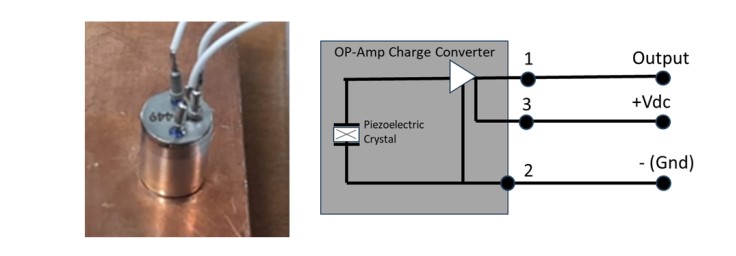}
    \caption{Piezoelectric detector glued to a copper metal slab. On the right a scheme of the detector unit and its electrical configuration.}
    \label{fig:piezo}
\end{figure}

When a particle crosses the sensitive volume of the detector, two main interaction mechanisms may occur:

\begin{itemize}
    \item \textbf{Ionization of the scintillator}: Charged particles lose energy via ionization processes within the plastic scintillator material. The resulting de-excitation leads to the emission of optical photons (scintillation light). 
    
    \item \textbf{Momentum transfer to metallic planes}: Charged or neutral massive particles, such as hypothetical nuclearites or strangelets, may deposit energy mechanically by inducing vibrations in the metallic copper plates. These vibrations are detected via piezoelectric accelerometers (Figure \ref{fig:piezo}).
\end{itemize}

\subsection{Detectors}

\begin{itemize}
      \item \textbf{Plastic scintillator planes:} Each of the four scintillator layers is coupled to Silicon Photomultipliers (SiPMs) that detect the signal generated by charged particles through ionization.
      Signal amplitude is proportional to electric charge of the incident particle.
      The scintillators used are of type EJ-208, a plastic material based on polyvinyltoluene, with the following properties:
     \begin{itemize}
          \item Decay time: 1.4~ns
          \item Rise time: 0.5~ns
          \item Emission wavelength peak: 391~nm
     \end{itemize}
        
In a future configuration, each scintillator will be divided into five strips of 2~cm~$\times$~10~cm to improve timing resolution.
The SiPMs (Silicon Photomultipliers) are directly coupled to the scintillators using optical grease and are connected to a Hamamatsu 13365-3050 module, which provides (Figure \ref{fig: saponetta}):
    \begin{itemize}
        \item Signal preamplification
        \item High-voltage supply (70~V)
        \item Temperature compensation to maintain stable gain
       \end{itemize}

\begin{figure}[H]
    \centering
    \includegraphics[width=0.5\linewidth]{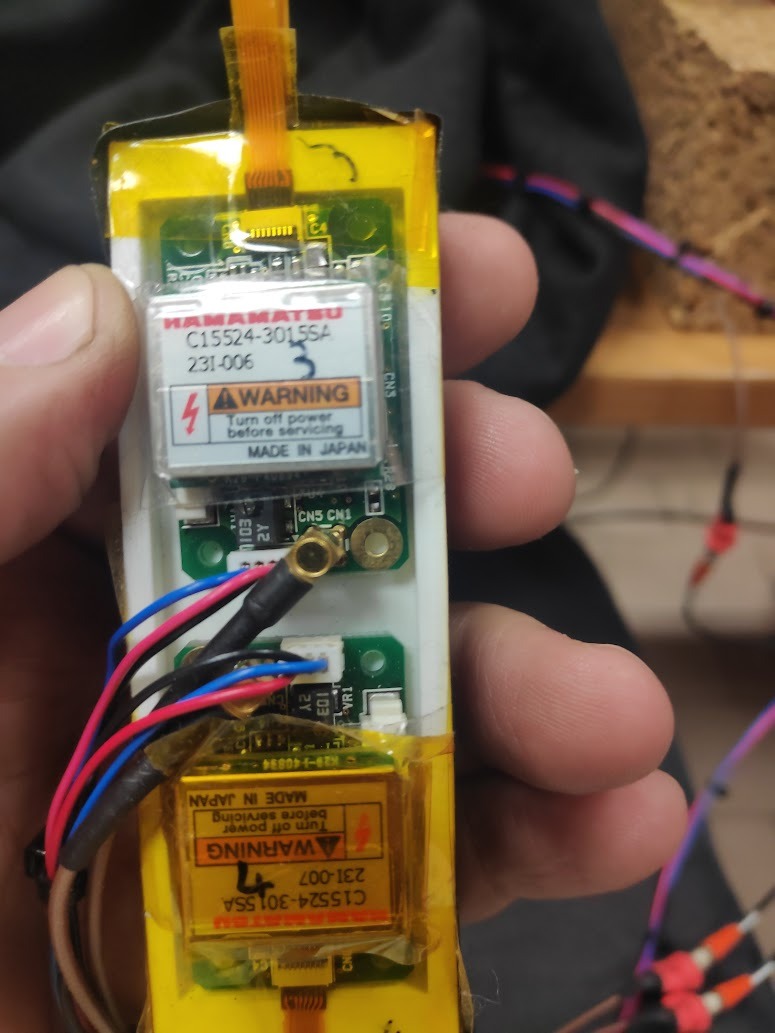}
    \caption{Scintillator bar used in the laboratory test setup. 
    The scintillator (10 cm long) is read at the two extremities by the Silicon Photomultipliers connected via flex cables to the two front-end electronics boards (on top in the figure). The two Hamamatsu C15524 DC-DC converters provide high voltage power supply and temperature gain compensation. 
    }
    \label{fig: saponetta}
\end{figure}

      \item \textbf{Copper planes with piezoelectric sensors:} Each of the four 
      metal plates are made of copper and serve as the support for the piezoelectric sensors, which detect vibrations induced by the passage of massive particles.
      Vibrations on the metal plate are transmitted to the crystal, producing an electric signal which is pre-amplified inside the unit. The piezoelectric detectors used are DSPM 160.1-V6V-1 units, with the following characteristics:
      \begin{itemize}
         \item Sensitivity: 100~mV/g $\pm$20\% ( 100~mV/(9.8m/s$^2$) $\pm$20\%), assuming a discernible signal with 5mV implies a maximum acceleration of $\simeq 0.5 m/s^2$. 
         \item Three-wire configuration for stable signal readout
          \item Integrated preamplifier to improve the signal-to-noise ratio
       \end{itemize}

As in the case of the scintillators, each plate is divided into five strips measuring 2~cm~$\times$~10~cm, with two piezoelectric detectors per strip (one on each side), for a total of 10 sensors per plane and 30 in total.
\end{itemize}
    
\section{Sensitivity of the SQM-ISS Detector to Primordial Black Holes}

We now consider the possibility of a PBH traversing the detector. The calculation of its ability to interact with atomic nuclei is analogous to that for SQM particles, but in this case the relevant scale is the Schwarzschild radius \( R_s \), which depends only on the PBH mass \( M \):

\begin{equation}
R_s = \frac{2GM}{c^2}
\end{equation}

If \( R_s \ll d_a \), where \( d_a \) is the atomic spacing of the detector material, the PBH will pass through the detector with negligible direct interaction with atomic nuclei. However, if the Schwarzschild radius is comparable to or larger than the atomic scale, the PBH will begin to interact directly with the nuclei—either by absorption or gravitational disruption along its path.

We take the atomic spacing of copper as a reference value:

\begin{align}
d_a &= 2.55 \times 10^{-10} \, \mathrm{m} \\
r_a &= \frac{1}{2} d_a = 1.27 \times 10^{-10} \, \mathrm{m}
\end{align}

By equating this to the Schwarzschild radius, we obtain the minimum PBH mass required for such direct interaction:

\begin{align}
r_a &= \frac{2GM_{\mathrm{min}}}{c^2} \\
M_{\mathrm{min}} &= \frac{c^2 r_a}{2G} \approx 8.55 \times 10^{16} \, \mathrm{kg}
\end{align}

Therefore, under this mechanism, only PBHs with masses greater than \( M_{\mathrm{min}} \) will have Schwarzschild radii large enough to interact effectively with the atomic lattice. In the intermediate case, where \( R_s \lesssim d_a \), partial interactions may occur depending on the lattice orientation and the PBH trajectory.

\subsection*{Gravitational Force Exerted by a PBH on a Disk:}

The \textsc{SQM-ISS} detector is also sensitive to the passage of primordial black holes (\textsc{PBHs})  via the reciprocal gravitational attraction. In this context, we consider the case in which a \textsc{PBH}, moving at typical galactic velocities ($v \sim 250~\mathrm{km/s}$), crosses the detector without directly interacting with matter, but instead induces an acceleration and displacement in the metallic planes through the gravitational field it generates. This can produce a piezoelectric signal even in the absence of ionization or other interaction mechanisms. For simplicity, the sensor plane is modeled as a thin disk of mass $M$ and radius $R$, lying in the plane $z = 0$, while the \textsc{PBH} is treated as a point mass moving along the $z$-axis, perpendicular to the disk. A numerical code was developed in \texttt{Python} to solve the motion of the \textsc{PBH} under the influence of the gravitational field generated by the disk and to compute the temporal evolution of its position, velocity, and acceleration. The code identifies the exact time at which the object crosses the plane ($z = 0$), produces plots of the kinematic quantities as functions of time, and analyzes in detail the acceleration profile near the disk ($|z| < 0.01~\mathrm{m}$), where the piezoelectric sensor is located. This model allows us to evaluate whether the induced acceleration on the disk exceeds the sensitivity threshold of the piezoelectric sensor and, consequently, whether the passage of the \textsc{PBH} can produce a detectable signal. It also provides a useful tool to estimate the minimum detectable mass of a \textsc{PBH}, and more generally, to characterize the response of the detector to purely gravitational events in the absence of direct matter interactions.

\subsection*{Gravitational interaction model}

We consider a thin disk of total mass \( M \), radius \( R \),
lying in the plane \( z = 0 \), and a PBH of mass \( m_{bh} \) located along the \( z \)-axis at a distance \( z \) from the center of the disk.
The surface mass density of the disk is:
\[ \sigma = \frac{M}{\pi R^2} \]
We work in polar coordinates \( (r, \theta) \) on the disk. A generic mass element of the disk is:
\[
d{m} = \sigma\, r\, d{r}\, d{\theta}
\]
The distance between the PBH and the element \( d{m} \) is:
\[
\ell = \sqrt{r^2 + z^2}
\]
The infinitesimal gravitational force \( d{\vec{F}} \) that the PBH exerts on the element \( d{m} \) is:
\[
d{\vec{F}} = \frac{G m_{bh} d{m}}{\ell^2} \cdot \frac{\vec{\ell}}{\ell} = \frac{G m_{bh} d{m}}{(r^2 + z^2)^{3/2}} ( -r\,\hat{r} + z\,\hat{z} )
\]
Only the \( z \)-component survives:
\[
d{F}_z = \frac{G m_{bh} d{m}}{(r^2 + z^2)^{3/2}} z = G m_{bh} z \frac{\sigma\, r\, d{r}\, d{\theta}}{(r^2 + z^2)^{3/2}}
\]

Integrating over the full disk:
\[
F_z = \int_0^{2\pi} d{\theta} \int_0^R \frac{G m_{bh} \sigma z r}{(r^2 + z^2)^{3/2}} d{r}
= 2\pi G m \sigma \int_0^R \frac{z r}{(r^2 + z^2)^{3/2}} d{r}
\]

\[
F_z = \pi G m_{bh} \sigma z \left[ -\frac{2}{\sqrt{u}} \right]_{z^2}^{R^2 + z^2}
= \pi G M_2 \sigma z \left( \frac{2}{z} - \frac{2}{\sqrt{R^2 + z^2}} \right)
\]

Simplifying:

\[
F_z = 2 \pi G m_{bh} \sigma \left( 1 - \frac{z}{\sqrt{R^2 + z^2}} \right)
\]

Substituting \( \sigma = \frac{M}{\pi R^2} \), we obtain the final expression:

\[
F_z = \frac{2 G m_{bh} M}{R^2} \left( 1 - \frac{z}{\sqrt{R^2 + z^2}} \right)
\]

The average acceleration imparted to the disk is:

\[
a(z) = \frac{F_z}{M} = -\mathrm{sign}(z)\frac{2 G m_{bh}}{R^2} \left( 1 - \frac{z}{\sqrt{R^2 + z^2}} \right)
\]
The motion of the disk along the \( z \)-axis can be modeled by a second-order ordinary differential equation:

\[
\begin{cases}
\dot{z} = v \\
\dot{v} = g(z)
\end{cases}
\]

where \( g(z) = a(z) \) is the gravitational acceleration due to the interaction with the disk. The initial conditions are:

\[
z(0) = d_0, \quad v(0) = v_0
\]

\subsection*{Minimum detectable acceleration from the piezoelectric}

We now calculate the sensitivity to PBH under the  assumption that it interacts only gravitationally with the piezoelectric detector, accelerating and displacing it due to its mass.   

The piezoelectric sensor employed in the SQM-ISS detector has a nominal sensitivity of
\[
S = 100~\mathrm{mV/g}
\]
where \(g\) is the standard gravitational acceleration. This implies that an acceleration of \(1~g\approx9.8~\mathrm{m/s^2}\) produces an output signal of 100 mV.
Assuming a linear response, the minimum detectable acceleration can be estimated from the minimum resolvable voltage signal.

A conservative threshold for detection can be set at 5mV, corresponding to an acceleration of 

\[
a_\mathrm{min} = \frac{5~\mathrm{mV}}{S} = \frac{5~\mathrm{mV}}{100~\mathrm{mV}/g} = 0.05~g \approx 0.49~\mathrm{m/s^2}.
\]

This value is the smallest acceleration expected to produce a measurable signal above the noise level, provided the signal duration is sufficient to ensure an adequate signal-to-noise ratio.
It therefore constitutes the lower bound for the acceleration induced on the disk that can be effectively detected by the piezoelectric system.

\begin{figure}[H]
    \centering
    \includegraphics[width=0.9\linewidth]   {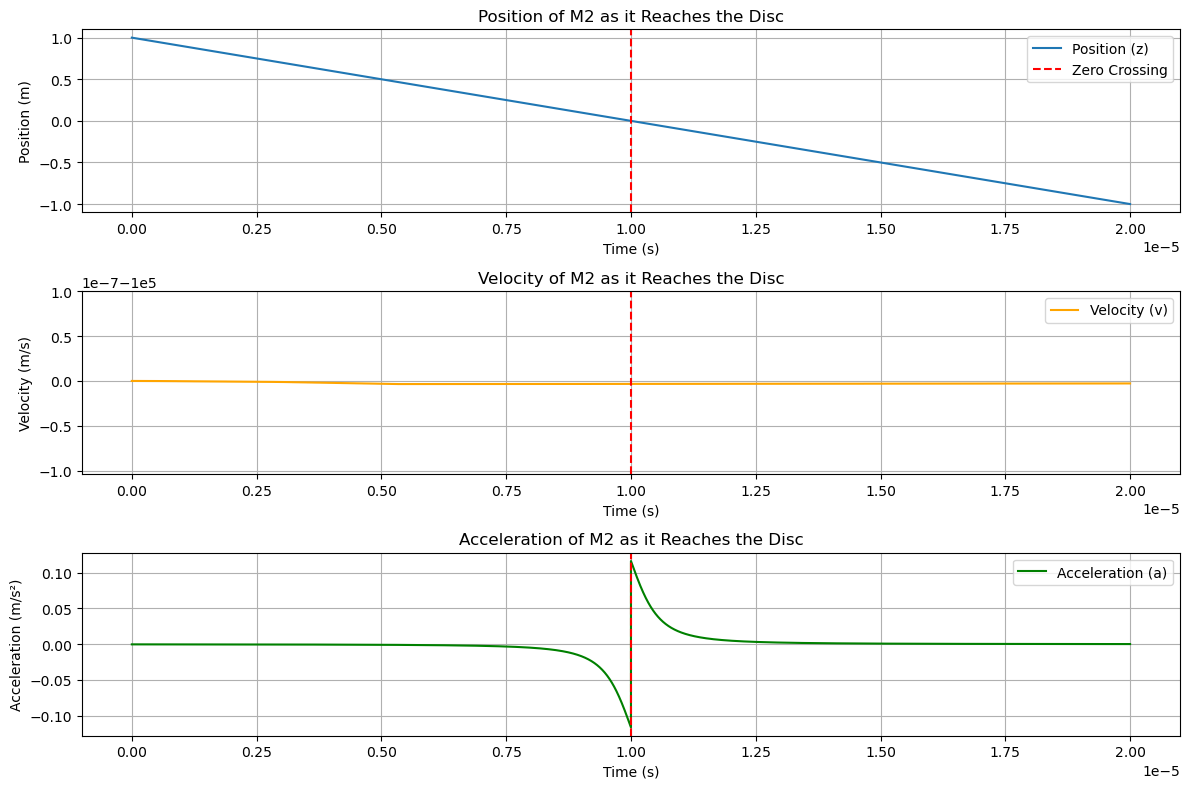}
        \includegraphics[width=0.9\linewidth]
    {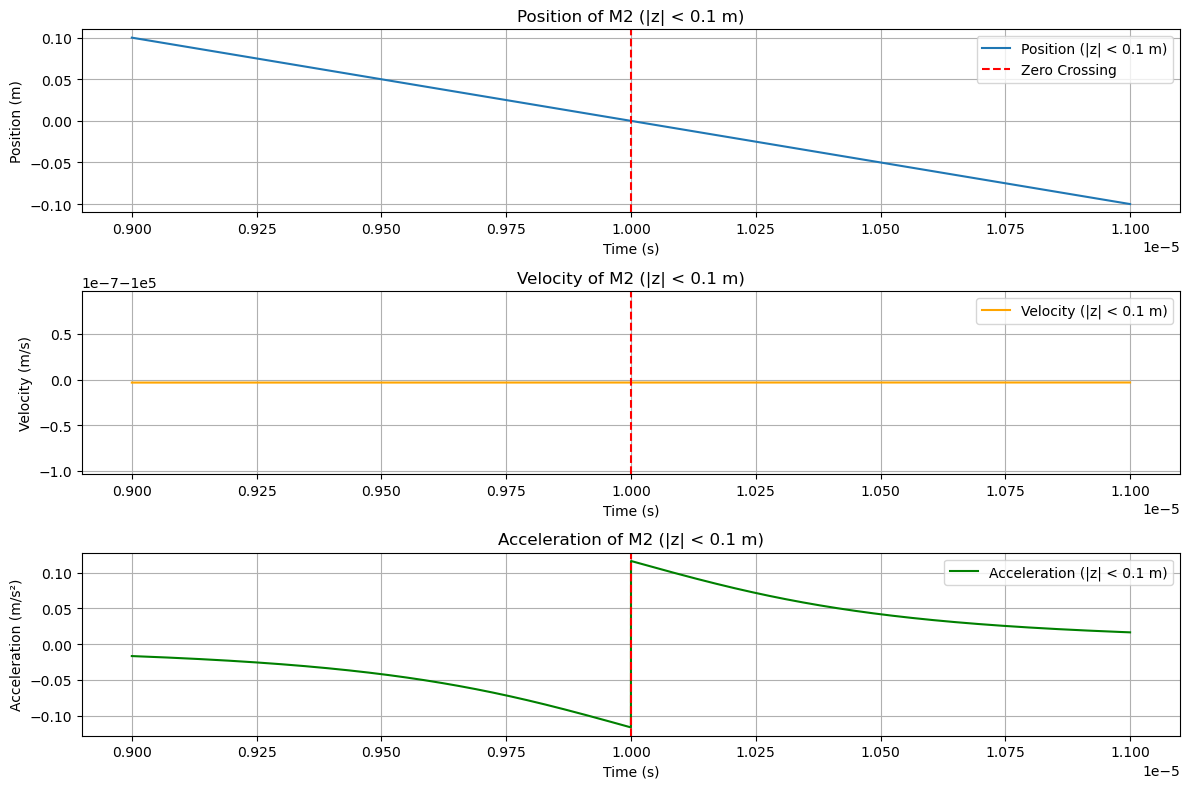}
    \caption{Simulated position, velocity, and acceleration of a PBH with mass \( m = 10^6~\mathrm{kg} \) moving along the \( z \)-axis with initial velocity \( v_0 = -10^5~\mathrm{m/s} \), representative of typical galactic velocities (left panels). The PBH is modeled as a test mass which generates a gravitational field capable of moving a thin massive disk located at \( z=0 \). This produces an acceleration profile as the black hole drags to itself the disc. The maximum acceleration is when the BH is close to the surface of the disc, before it penetrates it (not shown in the picture). The acceleration then decreases as the BH gets closer to the centre of the disc and then increases in the opposite direction. The right panels zoom  on a narrow region near the disk (\( |z| < 0.01~\mathrm{m} \)) to highlight the local kinematic behavior. The velocity remains nearly constant due to the high initial energy, while the acceleration exhibits a clear sign inversion across the disk plane. In this case the  simulation is maximum  approximately \( a_{\text{peak}} \simeq 0.01~g \approx 0.098~\mathrm{m/s^2} \), close to the minimum detection threshold of the piezoelectric sensor. This simulation was performed with a previous threshold estimate of \( a_\mathrm{min} = 0.01~g \), but the results remain valid for \( 0.05~g \) due to the linearity of the system. Therefore, under the assumption of a purely gravitational interaction, a PBH of \( 10^6~\mathrm{kg} \) represents the minimum detectable mass for the SQM-ISS system in its current configuration.
    }
    \label{fig:enter-label}
\end{figure}

\subsection*{Minimum detectable PBH mass}

To determine whether a passing PBH can produce a measurable signal in the detector, and establish a value for the minimum detectable mass, we consider the gravitational acceleration \( a(z) \) that it induces on the disk, as previously derived:
\[
a(z) = -\mathrm{sign}(z) \frac{2 G m}{R^2} \left(1 - \frac{|z|}{\sqrt{z^2 + R^2}} \right)
\]
A signal is detectable if the induced acceleration exceeds the sensor's minimum threshold \( a_{\text{min}} \), which defines the following condition:
\[
a(z) \geq a_{\text{min}}
\]
 
As discussed above, we can assume the minimum mass for a PBH when the maximum acceleration is about 0.01 g.



As shown in Figure \ref{fig:enter-label}, this corresponds to a  PBH with mass of \( 10^6~\mathrm{kg} \).

This first calculation shows that the highest sensitivity to Black Holes comes from the gravitational force exerted on the metal plane, rather than on the interaction with the atoms closest to its passage. This result will have therefore an impact on the detector design, with the metal planes capable of a slight vertical movement on the z (detector stack) axis, in order to let the accelerometer sense the displacement of the passage of the Black Hole. 


\reftitle{References}


\end{document}